\documentclass[epsf]{mn2e}
\def\gsim{\;\lower.6ex\hbox{$\sim$}\kern-7.75pt\raise.65ex\hbox{$>$}\;}
\def\lsim{\;\lower.6ex\hbox{$\sim$}\kern-7.75pt\raise.65ex\hbox{$<$}\;}

\title[CM Leo]{ANOMALOUS RR LYRAE STARS(?). III. CM LEONIS
\thanks{
E-mail:
gisella@bo.astro.it (GC), angela@bo.astro.it (AB), merighi@bo.astro.it (RM),
tosi@bo.astro.it (MT),
difabrizio@tng.iac.es (LDF), ditomaso@tng.iac.es (SDT),
marcella@na.astro.it (MM),
carretta@pd.astro.it (EC),
iivans@astro.as.utexas.edu (III),
chris@verdi.as.utexas.edu (CS),
smith@saucer.pa.msu.edu (HAS)
}}

\author[Di Fabrizio et al.]{L. Di Fabrizio$^{1,2}$, G. Clementini$^{1}$,
M. Marconi$^{3}$, E. Carretta$^{4}$, I.I. Ivans$^{5}$,
\newauthor
A. Bragaglia$^{1}$, S. Di Tomaso$^{2}$, R. Merighi$^{1}$, 
H.A. Smith$^{6}$, C. Sneden$^{5}$, M. Tosi$^{1}$\\
  $^1$ INAF-Osservatorio Astronomico di Bologna, Via Ranzani 1, I-40127 Bologna,
       Italy, \\
  $^2$ Centro Galileo Galilei \& Telescopio Nazionale Galileo,
       PO Box 565, 38700 Santa Cruz de La Palma, Spain,
       \\
  $^3$ INAF-Osservatorio Astronomico di Napoli, Via Moiariello 16,
        I-80131 Napoli, Italy,\\
  $^4$ INAF-Osservatorio Astronomico di Padova, Vicolo Osservatorio 5,   
       I-35122 Padova, Italy,
       \\
  $^5$ Department of Astronomy \& McDonald Observatory, University of Texas 
  at Austin, Austin, TX 78712, USA, 
  \\
  $^6$ Department of Physics \& Astronomy, Michigan State University, 
       East Lansing, Michigan 48824, USA}

\date{}

\begin{document}
\maketitle

\begin{abstract}
Time series of {\it B,V,I} CCD photometry and radial velocity  measurements
from high resolution  spectroscopy (R=30,000) covering the full pulsation cycle
are presented for the  field RR Lyrae star CM Leonis. The photometric data span
a 6 year interval  from 1994 to 1999,  and allow us to firmly establish the
pulsation mode and periodicity of the  variable.   The derived period {\it
P}=0.361699 days  ($\pm 0.000001$) is very close to the value published   in
the Fourth Edition of the General Catalogue of Variable Stars ({\it
P}=0.361732 days). However, contrary to what was previously found,  the
amplitude and shape of the light curve qualify CM Leo as a very regular first
overtone pulsator  with a prominent  hump on the rising branch of its
multicolour light curves. According to an  abundace analysis performed on three
spectra taken near minimum light (0.42 $< \phi <$ 0.61), CM Leo is  a
metal-poor star  with metal abundance [Fe/H]=$-1.93 \pm 0.20$. The photometric
and radial  velocity curves of CM Leo have been compared with the predictions
of suitable pulsational  models to infer tight  constraints on the  stellar
mass, effective temperature, and distance modulus of the star.  We derive a
true distance modulus of CM Leo of $\mu_0$=13.11 $\pm$0.02 mag and a  corresponding 
absolute magnitude of $M_V$=0.47$\pm$0.04.  This absolute magnitude, once 
corrected
for  evolutionary and metallicity  effects, leads to a true distance modulus of
the Large Magellanic Cloud of  $\mu_0$=18.43$\pm$ 0.06 mag, in better agreement with the
{\it long} astronomical distance scale.

\end{abstract}


\section{Introduction}

This is the third in a series of papers dealing with the detailed study of  the
pulsational  properties of a sample of field RR Lyrae stars that were known to 
exhibit anomalous  scatter and variations in  the shape and amplitude of their
light curves. In Clementini et al. (1995b; hereinafter Paper I) we  reported
results  from a first photometric study of a sample  of 8 field RR Lyrae stars
which, according to the literature, were classified  as fundamental mode
pulsators (RRab) with short periods and located at large  heights {\it z } from
the Galactic plane (Castellani, Maceroni \& Tosi 1983).  The circumstance of being far from the disc but with a
period that for RRab usually corresponds to roughly solar metallicity made
these stars fairly  anomalous. The photometry presented in Paper I 
demonstrated however that  five out of the eight variables were more likely to
pulsate in the first  overtone mode, and that, therefore, the short period
should not be taken  as indicative of high metallicity. Furthermore, it  showed
that some of the stars in the sample (namely CM Leonis, CU, BS  and BE Comae)
exhibited irregularities in their light curves. In order  to investigate  the
actual nature of these anomalies new observing campaigns were conducted  on
these four  stars. In Clementini et al. (2000, Paper II) we  reported the 
discovery that CU Comae is a {\it double-mode} RR Lyrae, the sixth detected in
the field of our Galaxy and the most metal-poor ever found. The present paper
is  devoted to  the discussion of the results obtained for CM Leonis (CM Leo)
from the  combination of its Paper I photometry with new data taken from 1995
to 1999,  and from   high resolution spectroscopy obtained in 1999. The
observations and
data sets are presented in Section 2. In Section 3 we 
discuss the results of  the analysis of the complete photometric data set.
In Section 4 we report the results  of our spectroscopic analyses (derivation
of a radial  velocity curve over the full pulsation cycle and  elemental
abundance analysis of the spectra of CM Leo  taken near minimum light).  In
section 5, we present results of the modeling of  the observed light and radial
velocity curves of CM Leo,  based on nonlinear pulsation models, computed with
the same physical and numerical assumptions as in Bono, Castellani \& Marconi
(2000), and derive an estimate of  the stellar mass, effective temperature, 
and distance modulus of the star. Finally, in Section 6 we summarize the main
derived quantities of CM Leo and  discuss the absolute magnitude we derive for
the star in the framework of the  {\it short} and {\it long} distance scale
dichotomy.

\section{Observations and reductions}

According to the fourth edition of the General Catalogue of Variable Stars 
(Kholopov et al. 1985, GCVS4) CM Leo ($\alpha_{2000}=11^h 56^m 14.3^s$, 
$\delta_{2000}=21^o 15' 30.2''$) is a fundamental mode RR Lyrae star (RRab) 
with a  rather short pulsation period P=0.$^d$361732 and 1.10 mag amplitude in
the  photographic light variation. Clementini et al. (1995b) obtained CCD {\it
V} photometry of  CM Leo in 1994.  With a period P=0.$^d$361479 they basically
confirmed the periodicity of  the GCVS4,  but showed that the light curve  has
a sinusoidal shape with amplitude  of 0.49 mag, and  appears to be split in two
well-separated subcurves with about 0.1 mag difference in  amplitude (see
figure 3a in Paper I).   Clementini et al. (1995b) concluded that CM Leo is a
{\it c}-type RR Lyrae with  irregularities in its light curve.  Three different
causes   were suggested to possibly explain the anomaly in the light curve, 
namely double-mode or non-radial mode pulsation (Olech et al. 1999), and the 
Blazhko effect (Blazhko 1907). Given the small number of data points (71)  and
the short time baseline (2 months), the 1994 observations were  insufficient to
pinpoint the actual source of the irregularities.  New photometry of the star
was obtained from 1995 to 1999 with the Loiano 1.52 m  telescope operated by
the Bologna Observatory, and the  60 cm  of the Michigan State University
(MSU). High resolution spectroscopy  of the star was also obtained in 1999,
with the 2.7 m telescope of the McDonald Observatory.  \subsection{Photometric
data} The new photometric data of CM Leo  consist of {\it B,V,I} CCD
observations in the Johnson-Cousins photometric system  defined by Landolt
(1983, 1992) standard stars,  obtained with the Loiano 1.52 m  telescope, on 27
nights from January 1995 to  February 1999. They were complemented by 171 and
30 V frames obtained with the  MSU 60 cm  telescope in 1998 and on two
consecutive nights in  February 1999, respectively. At both telescopes,  the
1998 campaign  consisted of several consecutive nights (from 5 to 10) in three
major runs about one month apart, to test both the possibility of double-mode
or non-radial mode pulsation  (which are both known to occur on cycle-to-cycle
timescales),  and of the Blazhko effect (whose periodic modulation of the light
curve  has typical timescales of tens of days). In addition, the 1999 campaign
consisted of two nights of coordinated observations at the  Loiano and MSU
telescopes,  which allowed us to obtain continuous  photometry of CM Leo over
10.4 consecutive hours. This interval is larger than  a full pulsation cycle of
CM Leo (8.7 hours) thus permitting us to further check  on the cycle-to-cycle
nature of the irregularities in the light curve.  On these same nights high
resolution spectroscopy  of the star was also secured with the 2.7 m telescope
of the McDonald Observatory  to obtain the radial velocity curve, and to
perform a chemical abundance analysis. The journal of the photometric
observations is given in Table~1.

\begin{table}
\begin{center}
\caption{Journal of the photometric observations}
\vspace*{5mm}
\begin{tabular}{c c c c c c}
\hline
\hline
\multicolumn{1}{c}{Year} &
\multicolumn{3}{c}{N.of Observations}&
\multicolumn{1}{c}{Observed intervals}&
\multicolumn{1}{c}{Telescope}\\
     &   $B$  &  $V$  &  $I$ & (HJD$-$2452400000)&   \\
\hline
1995 &  17  &  28 &  15  & 49730 - 49756& 1.52m \\
1996 &  21  &  19 &  19  & 50122 - 50131& 1.52m \\
1997 &  12  &  23 &  11  & 50507 - 50508& 1.52m \\
1998 &  55  & 315 &  63  & 50841 - 50910& 1.52m \\
1998 &      & 171 &      & 50845 - 50921& 0.6m \\
1999 &      &   8 &      & 51223& 1.52m \\
1999 &      &  30 &      & 51223 - 51224& 0.6m \\
Tot  & 105  & 594 & 108  &              &      \\
\hline
\end{tabular}
\end{center}
\label{t:tab1}
\end{table}

Observations at the 1.52 m telescope were obtained with BFOSC (Bologna Faint
Object Spectrograph \&  Camera) mounting a Thomson 1k$\times$1k CCD (0.5
arcsec/pixel, field of view 9.6 arcmin$^2$); filters in the 
Johnson-Cousins photometric system were used. The field  of CM Leo, taken from
a CCD image, is shown in Figure~1.  Observations at the 60 cm MSU telescope 
were obtained with an Apogee Ap7 CCD,
providing an 11 arcmin$^2$ field of view.

\begin{figure*}
\vspace{12cm}
\includegraphics{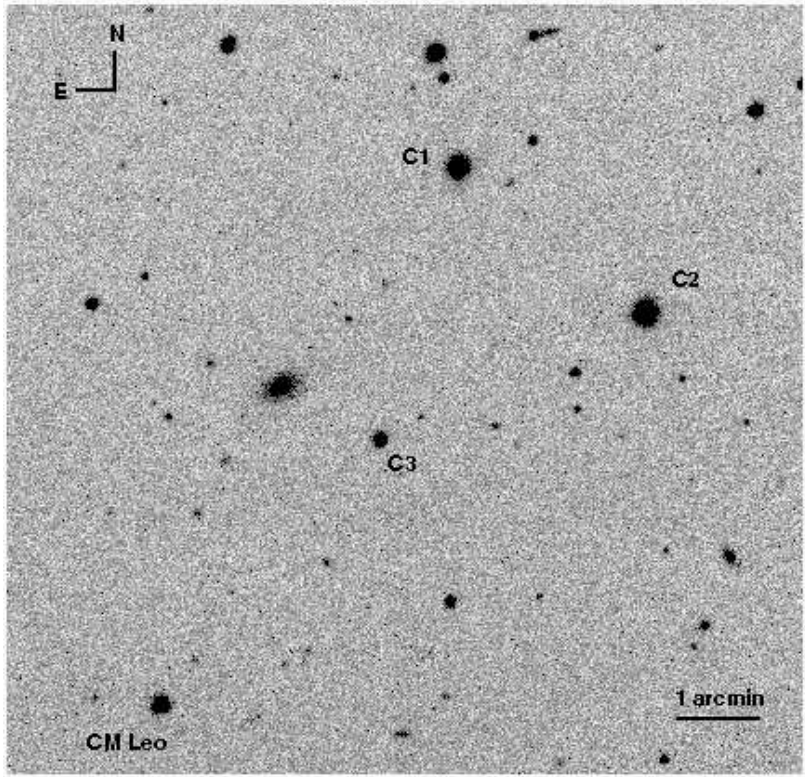}
\caption{A 9.6 arcmin$^2$ CCD image of the CM Leo field. The other
three stars marked were used as comparison stars to obtain differential
measurements of the magnitude variation of CM Leo. }
\end{figure*}

Three other objects are marked in Figure~1: the comparison stars C1, C2  and
C3.   Since only a small fraction of the observations of CM Leo were taken in 
truly photometric conditions, the light curve of the variable  has been derived
in terms of differential  magnitude with respect to these comparison stars. 
 Data were pre-reduced,  and
instrumental magnitudes of the variable and comparison stars were  derived by
direct photon counting  using standard routines for aperture  photometry in
IRAF\footnote{
IRAF is distributed by the National Optical  Astronomy Observatories, which is
operated by the Association of Universities  for Research in Astronomy, Inc.,
under cooperative agreement with the  National Science Foundation.}. 
The photometric data were tied to the standard photometric system through
observations of 35 standard stars selected from  Landolt (1983, 1992; for
details on the  calibration procedure see Di Fabrizio 1998). 

\subsubsection{The comparison stars}

The choice of the best comparison star turned out to be a crucial point  in the
study of the variability of CM Leo. Of the three isolated stars falling in the
CCD, C1 and C2 are of comparable  luminosity  ($V \sim$12.2-12.5 mag), while C3
is about 2.5 mag fainter and has a redder  colour  ($B-V \sim 1.0$ mag).  C1
was used as comparison star in Paper I. Its stability was checked at the time
against star C2.  However, some doubts about the constancy  of both  C1 and C2
were present, due to small trends in the  instrumental $V_{C1}-V_{C2}~ {\it
versus}$  heliocentric Julian day (HJD)   plots, leading to magnitude
differences of the order of some hundredths of a  magnitude from one run to
the  other. The significantly increased number of observations and the 
homogeneity and  accuracy of the data reduction procedure of the present  study
have allowed us to look into this matter in more detail.

\begin{figure*}
\vspace{16cm}
\includegraphics{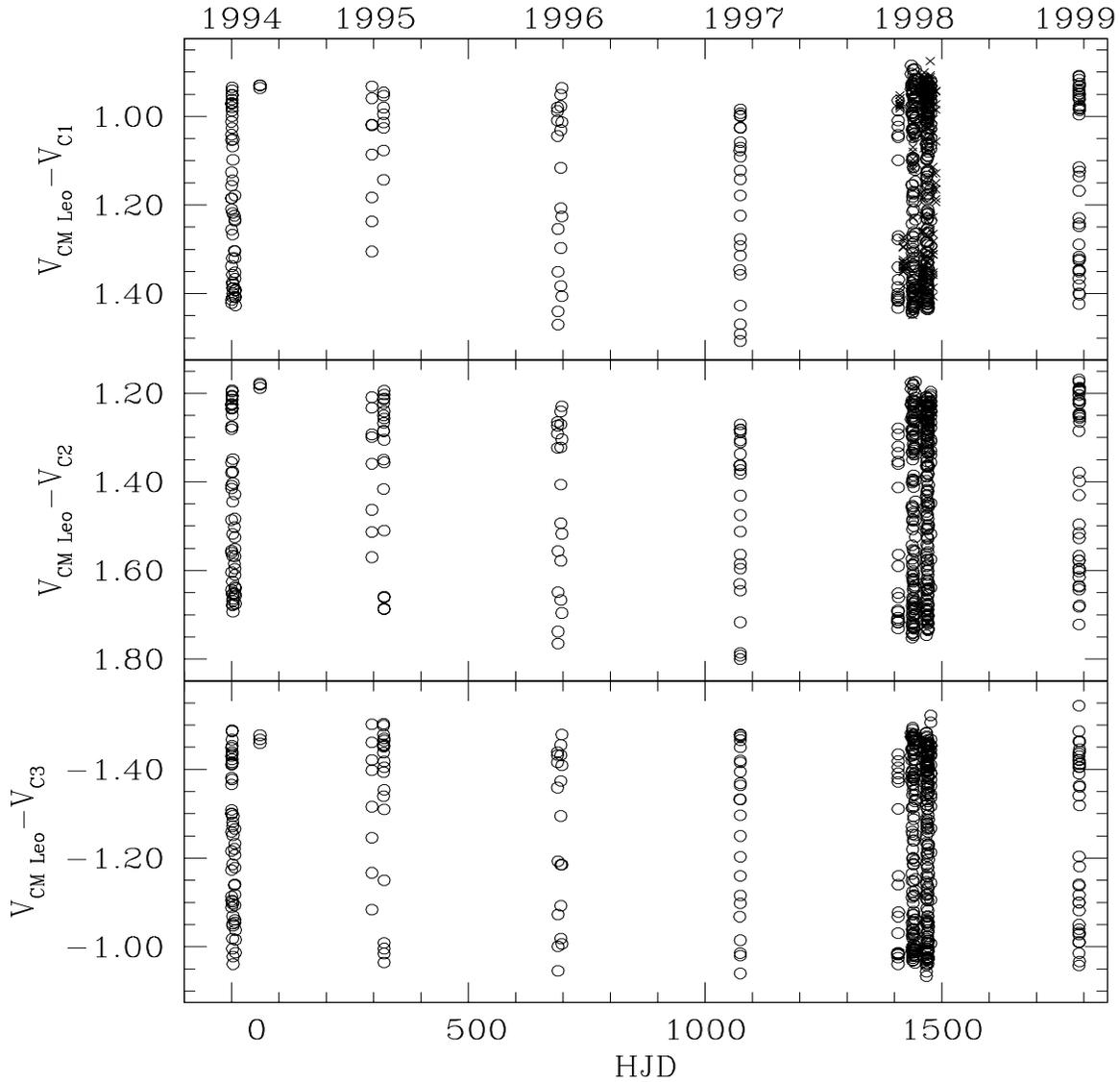} 
\caption{Differential instrumental $V$ magnitudes of CM Leo with respect to 
star C1, C2 and C3. The zero point of the x axis is the HJD of the first 
photometric observation in 1994 (HJD=2449434.376).}
\end{figure*}

Figure~2 plots the differential instrumental $V$ magnitudes of CM Leo  with
respect to the comparison stars C1, C2 and C3 (top, middle and lower  panels,
respectively) as a function of the HJD of observation, for the 1994-1999
data-set.  The top panel includes also the 1998 MSU observations  (crosses, 171
data-points).  With only the exception of the 1995 CM Leo$-$C1 subset, in each
year of observation data shown in the three  panels cover the full light 
variation from maximum to minimum light. This permits to note a variation  of
about 0.05-0.1 mag both in the minimum and maximum values  of  $V_{CM
Leo}-V_{C1}$ and $V_{CM Leo}-V_{C2}$.  No variation is instead present in the 
$V_{CM Leo}-V_{C3}$ {\it versus} HJD plot. Furthermore, Figure~3 and Figure~4
show   the differential instrumental $V$ and $I$ light curves of CM Leo  with
respect to the three comparison stars, with data folded according to the best
fit period of the  variable (P=0.361699, see Section~3).

\begin{figure*}
\vspace{15cm}
\includegraphics{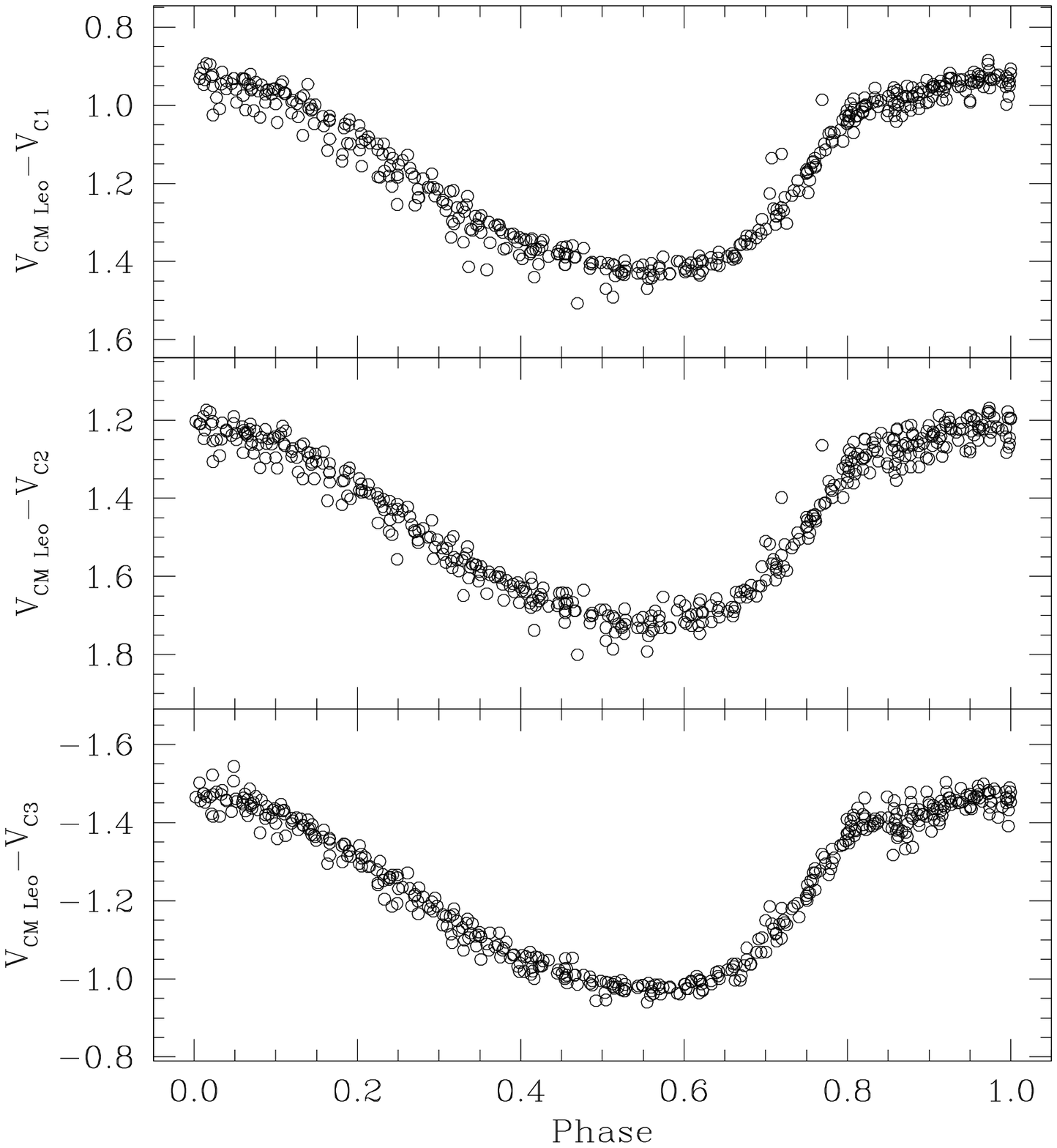}
\caption{Differential instrumental $V$ magnitudes of CM Leo with respect to
stars C1, C2 and C3, respectively.}
\end{figure*}

\begin{figure*}
\vspace{15cm}
\includegraphics{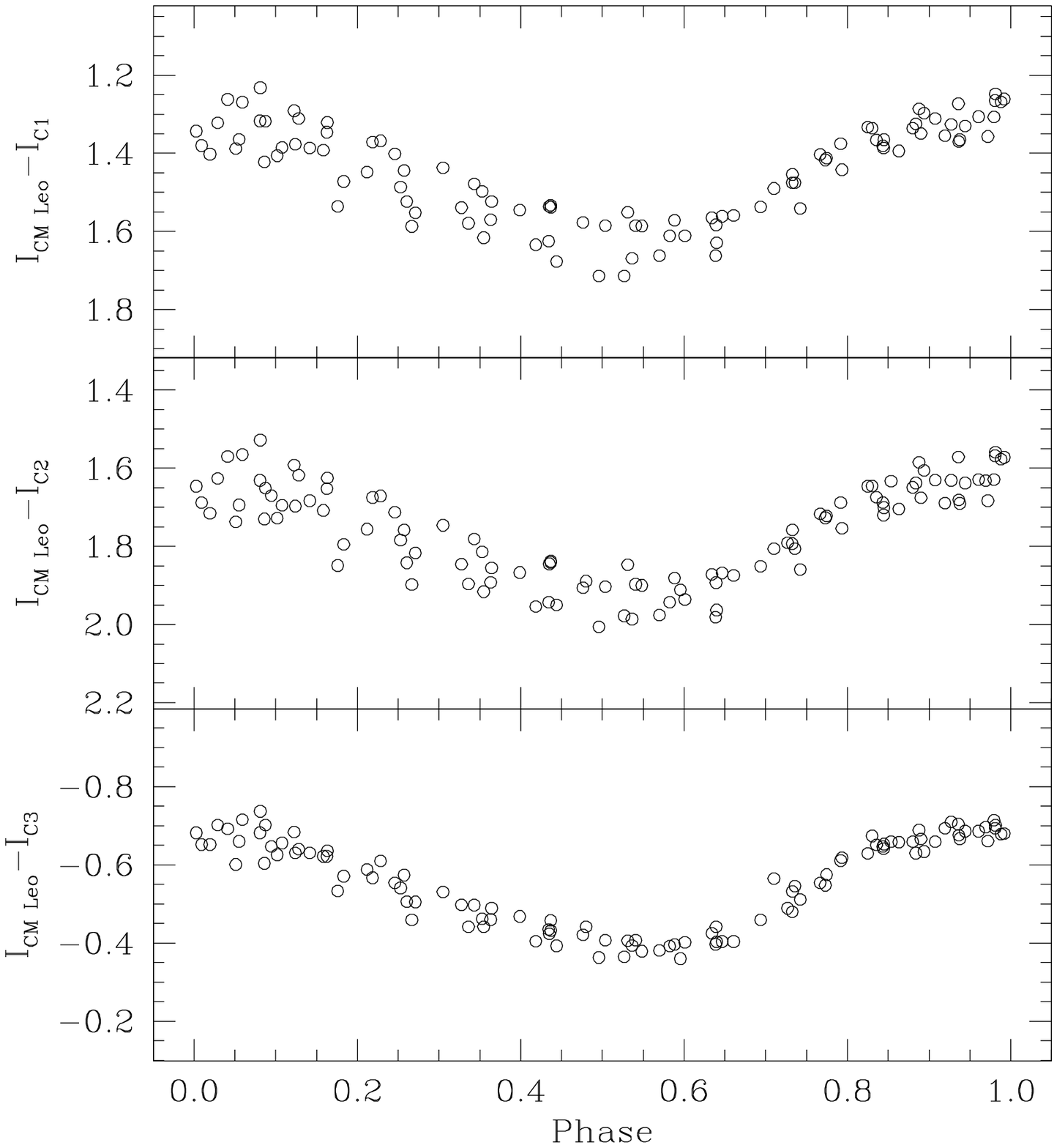}
\caption{Differential instrumental $I$  magnitudes of CM Leo with respect to
stars C1, C2 and C3, respectively.}
\end{figure*}

The scatter in the $V$ light curve of CM Leo, measured by the r.m.s. from the
best fit model,  is reduced from 0.03 mag when using C1 and C2 as comparison
stars,  to 0.02 mag when using C3. A similar effect is seen for the $I$ light
curve (scatter going from 0.05 down to 0.02 mag) while the effect is smaller 
in $B$. The anomaly found in Paper I ---  a splitting of the $V$ light curve 
in two well-separated subcurves with about 0.1 mag difference in  amplitude ---
is  thus solved, and details of the light curve variation of CM Leo, as for
instance  the hump on the ascending branch, are better defined  when star C3 is
used as reference. This suggests  that light variations may affect stars C1 and
C2, as confirmed by the  plots in Figure~5 where we show the instrumental
differential  $V$, $B$ and $I$ magnitudes of stars C1 and C2  with respect to
star C3,  {\it versus} the HJD of observation, of the entire data set.

\begin{figure*}
\vspace{15cm}
\includegraphics{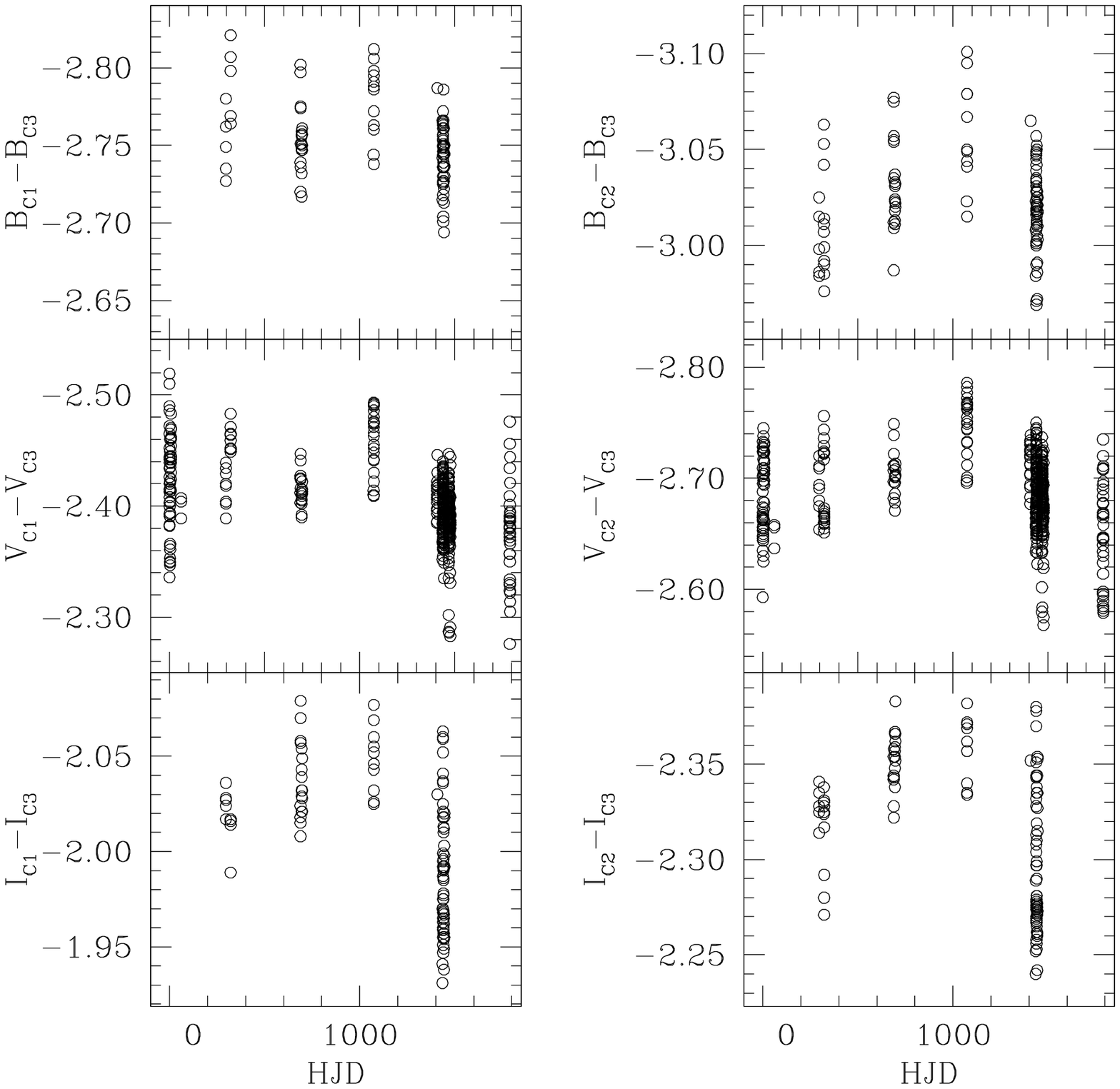}
\caption{Differential instrumental $V, B$, and $I$ magnitudes of star C1 (left
panels) and C2 (right panels) with respect to star C3.}
\end{figure*}

A variation of 0.05-0.20 mag is indeed present in these plots: the  variation
appears to be slightly larger for star C2 than for star C1, and in both cases
is stronger in $V$ and $I$ and less detectable in  $B$. The two stars may be
small amplitude, long period variable stars,  or they could have faint red
companions,  however  a conclusion on the actual cause of their
light variation cannot be reached from the  present data. In any case, we
conclude that  they cannot be used as references for CM Leo, whose light
variation is better studied by comparison with star C3. Coordinates and
calibrated magnitudes of the 3  stars in the field of CM Leo are given in
Table~2, where the  uncertainties include both  the internal error contribution 
(about $\pm$0.02 mag in all three passbands) and the systematic errors in going 
to the standard system ($\pm$0.02 mag in 
$V$ and $B$, and $\pm$0.03 mag in $I$). The slightly larger
 error of the $I$
measurements, in spite of C3 being brighter in $I$, is due to the larger
uncertainty of the $I$ absolute photometric calibration (fewer standard stars
were observed in I, see Di Fabrizio 1998).
The stars  are identified in column  2 of Table~2 by
their number on the second version of the {\it Hubble Space  Telescope} Guide
Star Catalogue (GSC2). Magnitudes given for C1 and C2  have been derived as  
average of all available differential measurements with respect to star C3, 
tied to the standard  system through the absolute calibration of star C3. 

\begin{table*}
\begin{center}
\caption{Magnitudes and coordinates of the 3 comparison stars in the field of 
CM Leo}
\begin{tabular}{c c c c c c c}
\hline
\hline
\multicolumn{1}{c}{Star} &
\multicolumn{1}{c}{N$_{GSC2}$}&
\multicolumn{1}{c}{$\alpha_{2000}$} &
\multicolumn{1}{c}{$\delta_{2000}$} &
\multicolumn{1}{c}{$V$} &
\multicolumn{1}{c}{$B$} &
\multicolumn{1}{c}{$I$} \\
\hline
C3 &N2001000143& 11 56 03.7& 21 18 27.9& 14.90 $\pm$ 0.03 & 15.87 
$\pm$ 0.03 & 13.87 $\pm$ 0.04 \\
\hline
C1 &N2001000131& 11 55 59.9& 21 21 31.0& 12.50 & 13.12 & 11.86 \\
C2 &N2001000136& 11 55 51.0& 21 19 53.5& 12.21 & 12.85 & 11.55 \\
\hline
\end{tabular}
\vskip 0.5 cm
\end{center}
\label{t:tab2}
Note. - C1 and C2 are suspected variable stars with amplitude of the light
variation of about 0.05-0.2 mag. Magnitudes given for them are 
the average of all the available differential values with 
 respect to star C3, tied to the 
standard system through the absolute calibration of star C3.
\end{table*}

The light variation of CM Leo was then studied from the differential photometry
with respect to star C3. Photometry published for CM Leo in Paper I used star
C1, so  the  1994 data have all been remeasured. Unfortunately, star C3 was not
sufficiently exposed in the 1998 MSU observations, so this photometry cannot be
used in the study of the period. The $B, V,$ and $I$ magnitudes of CM Leo
relative to the comparison star  C3  along with the Heliocentric Julian date
at  mid-exposure are listed in Table~3 (which will be provided in  its entirety
through the CDS). According to the errors quoted in Table~2 we estimate that 
the photometric accuracy of each individual data point is  $\pm 0.03$ mag in
$V$ and $B$, and $\pm 0.04$ mag in $I$, including both random and 
systematic uncertainties.  

\begin{table*}
\begin{center}
\caption{The photometric observations of CM Leo with respect to star C3.}
\vspace*{5mm}
\begin{tabular}{r c r c r c}
\hline
\hline
\multicolumn{1}{c}{~~HJD}&
\multicolumn{1}{c}{~$\Delta V$}&
\multicolumn{1}{c}{~~HJD}&
\multicolumn{1}{c}{~$\Delta B$}&
\multicolumn{1}{c}{~~HJD}&
\multicolumn{1}{c}{~$\Delta I$}\\
($-$2400000) & & ($-$2400000) & & ($-$2400000) &\\
\hline
49755.474538 &~~$-$1.438~~&49755.463750   &~~$-$2.092~~& 49755.482756
&~~$-$0.651~~\\
    0.490685 &~~$-$1.456~~&      0.502722 &~~$-$2.191~~&     0.522006
    &~~$-$0.686~~\\
    0.513602 &~~$-$1.503~~&      0.541092 &~~$-$2.241~~&     0.560607
    &~~$-$0.601~~\\
    0.530212 &~~$-$1.499~~&      0.579381 &~~$-$2.126~~&     0.599440
    &~~$-$0.622~~\\
    0.552146 &~~$-$1.479~~&      0.618828 &~~$-$1.962~~& & \\
    0.568652 &~~$-$1.451~~& & & & \\
    0.590343 &~~$-$1.394~~& & & & \\
    0.607762 &~~$-$1.340~~& & & & \\
\hline
\end{tabular}
\end{center}

\smallskip
Note. - Table 3, which also includes updated Clementini et al. (1995b) 
photometry, is presented in its entirety at CDS, and a portion is shown here
for guidance  regarding its form  and content.
\label{t:tab3}
\end{table*}

\subsection{Spectroscopic data}

CM Leo was observed with the ``2d-coud\'e'' spectrograph (Tull et al. 1995)
of the 2.7m telescope at the Mc Donald Observatory on UT 1999 February 14
and 15. 
The  spectrograph mounted the E2 echelle grating (R=63 000), and a 
2048$\times$2048 thinned, grade 1 Textronix CCD.								 
Seeing conditions during the nights varied from 1.5$^{\prime \prime}$ to 
1.7$^{\prime \prime}$.
Our total wavelength coverage extended from 3670~\AA\ to 9900~\AA.
We employed the data from 3924~\AA\ thorough 8263~\AA\ in our metallicity 
determinations and restricted the radial velocity determinations to the 
best-suited orders, in a region from 
about 4400~\AA\ to 5300~\AA.
Exposures of a Thorium-Argon lamp to obtain the wavelength calibration were
taken at beginning, middle and end of each night. The radial velocity standard
star  HD 58923 (RV=+17.8 km s$^{-1}$, Wilson 1953) was also observed, to be
used as a template for the cross-correlation measure of the radial
velocities  of CM Leo and to set the radial velocity zero point.

Given the relative faintness of our target ($<V> =13.66$ and $V_{min}  \sim$
13.9  mag) and the severe constraints on the exposure length, which we limited 
to 20-30 min in order to avoid  phase blurring on the pulsation cycle, we used
a 1010 $\mu$m slit (which projects to 2 arcsec on the sky) and a 2$\times$2
binning in order to increase the S/N ratio. We obtained 12 high-resolution
spectra of the star, covering a full pulsation cycle. The spectra have
resolution R=30,000 and moderate S/N ratio (S/N$\sim$ 50 for a single
spectrum).  The FWHM measured  from the ThAr comparison lamp lines is $\sim$
0.2~\AA\ (at $\lambda \sim$ 6300~\AA).  

Data were reduced using standard IRAF  processing tools  to subtract the bias
level, trim images, identify and correct ``bad''  pixels by means of long
exposure dark frames and  flat fields, and divide by a scaled flat field  image
(comprised of 20 spectra taken with a quartz lamp through a  blue-pass filter).
The minimum order separation  was 10 arcsec, leaving sufficient interorder
background free of  contamination from starlight in order to adequately remove
the (small) scattered light contribution.  Cosmic ray excision was  performed
using {\it lineclean}.  

The individual orders were  extracted from the two-dimensional  data to obtain
one-dimensional summed spectra of the stellar and comparison lamp frames.  
The dispersion relation to transform from pixel to wavelength was determined
from the ThAr exposures, and applied to all stellar spectra, interpolating in
time over shifts in the comparison  line positions on the chip through the
course of a night.  We  verified that the resulting wavelength solutions gave
correct  positions of known atmospheric emission and absorption features on 
each spectrum.  Geocentric and heliocentric Julian dates were  computed, and
later  used to compute the corrections to the observed  radial velocity due
to the diurnal, monthly, and yearly motions.

\section{Analysis of the photometric data: period search}

Period search was performed on the differential photometry of CM Leo with
respect  to the reference star C3 and the full 1994-1999 photometric data-set 
(94 $B$, 442 $V$ and 102 $I$ useful data points) using  GRATIS (GRaphycal
Analyzer TIme Series), a code developed at the Bologna Observatory  (see Di
Fabrizio 1998, Paper II).  We used the Lomb periodogram (Lomb 1976, Scargle
1982) on a wide period interval to derive a first guess of the periodicity, 
then a  truncated Fourier series algorithm (Barning 1963) to    refine the
period definition and  find the best fit model. The period search employed each
of the complete $B, V,$ and $I$ data-sets. Figure~6 shows the periodograms of
CM Leo $V, B,$ and $I$ differential  magnitudes with respect to star C3, 
obtained using the Lomb algorithm to identify the most  probable frequency of
the data on a wide period interval of 0.23-0.80 day. A well defined peak is
present in the plots at  $\omega$=2.76 (P$\sim$0.$^d$362),  with two lower
peaks at $\omega$=1.76 and $\omega$=3.76 respectively, which are aliases of the
actual periodicity.

\begin{table}
\begin{center}
\caption{The heliocentric radial velocities of CM Leo}
\vspace{5 mm}
\begin{tabular}{l r c c c}
\hline
\hline
\multicolumn{1}{c}{Spectrum} &
\multicolumn{1}{c}{~~HJD} &
\multicolumn{1}{c}{$\Phi$~~~} &
\multicolumn{1}{c}{${\rm RV}$~} &
\multicolumn{1}{c}{Error}\\
\multicolumn{1}{c}{} &
\multicolumn{1}{c}{} &
\multicolumn{1}{c}{} &
\multicolumn{1}{c}{km s$^{-1}$} &
\multicolumn{1}{c}{km s$^{-1}$}\\
\hline
~~~W1&51223.759995 & 0.227 & 22.83 & 1.59\\
~~~W2&    0.781506 & 0.286 & 25.40 & 1.90\\
~~~W3&    0.802912 & 0.345 & 27.96 & 1.18\\
~~~W4&    0.830009 & 0.420 & 31.04 & 1.22\\
~~~W5&    0.864014 & 0.514 & 33.65 & 2.12\\
~~~W6&    1.727197 & 0.901 & 11.04 & 2.99\\
~~~W7&    1.741795 & 0.941 & 12.38 & 3.07\\
~~~W8&    1.756334 & 0.981 & 10.42 & 2.57\\
~~~W9&    1.770790 & 0.021 & 11.32 & 2.42\\
~~~W10&   1.984161 & 0.611 & 35.91 & 1.90\\
~~~W11&   2.009543 & 0.681 & 36.91 & 1.41\\
~~~W12&   2.024329 & 0.722 & 32.72 & 2.79\\
\hline						      
\end{tabular}					      
\end{center}					      
\label{t:tab4}					      
\end{table}

\begin{figure} 
\vspace{16.5cm} 
\includegraphics{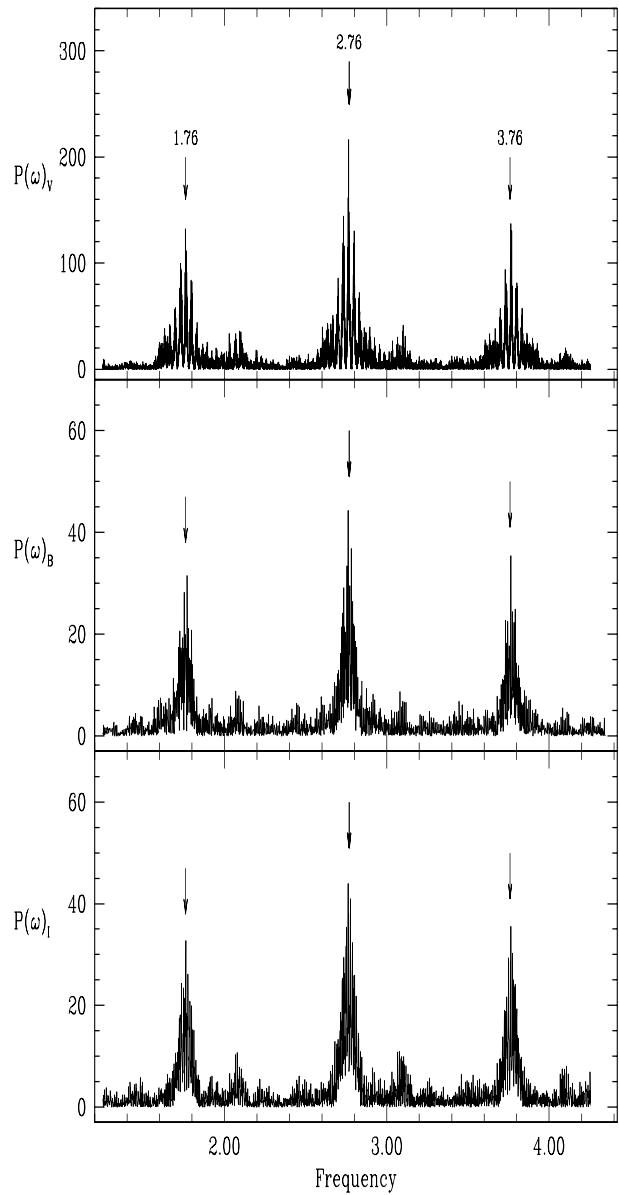}
\caption{Periodograms of the $V, B,$ and $I$ differential magnitudes of CM Leo
which identify the most probable frequency of the data ($\omega$=2.76), and
its two aliases at $\omega$=1.76 and $\omega$=3.76.}
\end{figure}

We then reduced the interval around the most likely periodicity using  the
Fourier algorithm to find the best fit. The  period obtained using a twelve
harmonics best-fitting Fourier series on the $V$ data, a 6 harmonics series on
the $B$ data, and a 5 harmonics series on the  $I$ data is
P=0.$^d$361699$\pm$0.000001 and the epoch of maximum light is
E=2450841.362285.  Calibrated light curves for CM Leo in $B, V,$ and $I$, based
on the full photometric data-set, and phased according to these ephemerides,
are shown in Figure~7. 

\begin{figure*}
\vspace{15cm}
\includegraphics{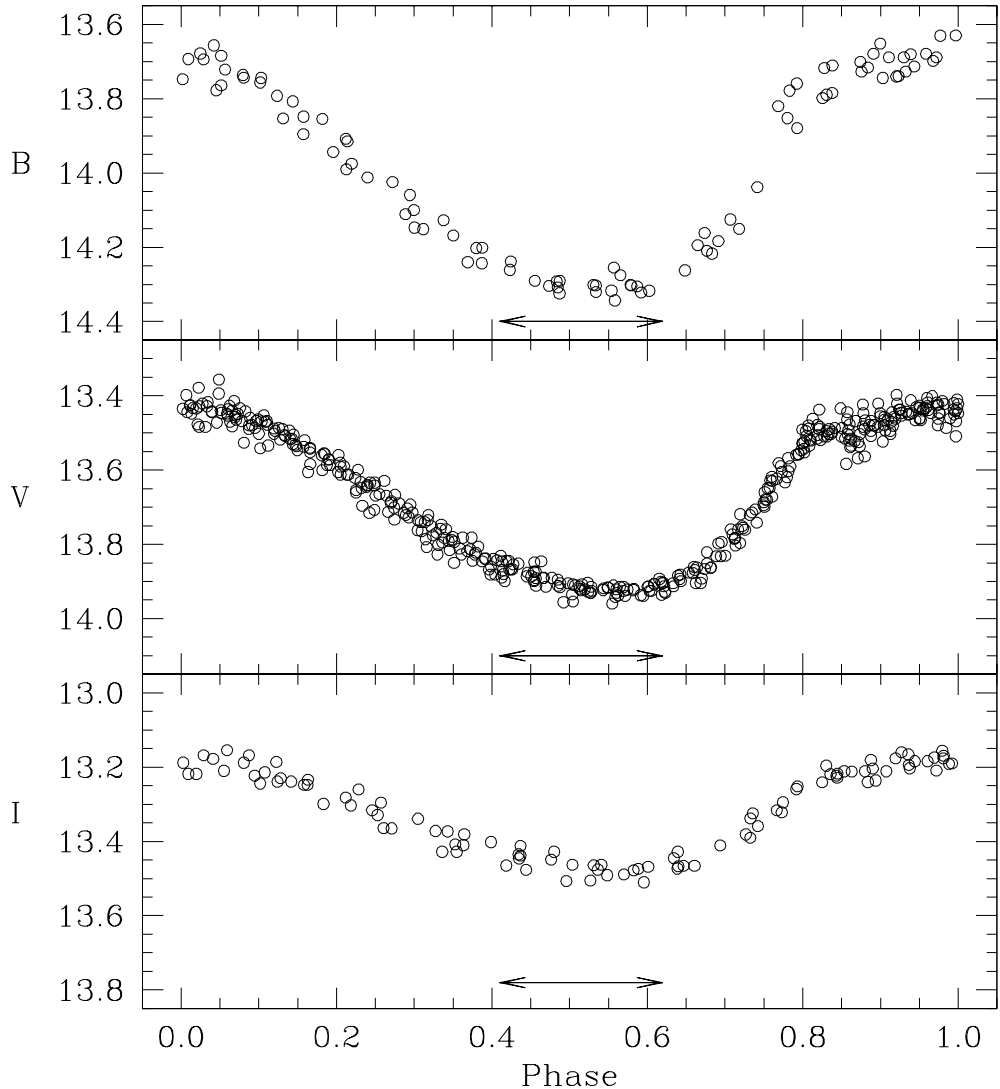}
\caption{Calibrated $B, V,$ and $I$ light curves of CM Leo. Data are phased
according to a period of pulsation of P=0.$^d$361699. The complete data-set
spanning 6 years of observations (1994-1999) is plotted. The bars  indicate the
phase interval covered by the high resolution spectra used to measure metal
abundances (see Section 4.2).}
\end{figure*}

The r.m.s. deviation  from the best-fitting model is $\pm 0.023$ mag in $V$,
$\pm 0.032$ mag in $B$  and $\pm 0.021$ mag in I. These residuals are well 
accounted for by random photometric uncertainties alone (about $\pm 0.02$ mag 
in 
all three photometric passbands, see Section 2.1.1).  The derived period 
agrees fairly well with that
published in the GCVS4 (P=0.$^d$361732). The amplitude of the light variation
is 0.498 mag in $V$,  0.631 in $B$, and  0.301 in $I$.  On the basis of these
amplitudes and the shape of the light curve,  CM Leo is classified  as a very
regular first overtone RR Lyrae star. We thus confirm Paper I  classification
of CM Leo as a {\it c}-type RR Lyrae, at variance  with the GCVS4 original
classification as an {\it ab}-type RR Lyr, with amplitude about twice the
present value.

\section{Analysis of the spectroscopic data}

\subsection{The radial velocity curve}

Radial velocities were measured from the wavelength calibrated spectra of CM
Leo using  a cross correlation technique ({\it fxcor} in IRAF). Twelve orders
containing weak metal lines (the best suited to measure radial velocities) in
the region from about 4400 to 5300 \AA~~ were cross-correlated against the same
orders in the spectra of the radial  velocity standard star HD 58923. Table~4
lists  the derived heliocentric radial velocities and corresponding errors.
Average precision of the radial velocity measurements  is about 2 km s$^{-1}$. 
The heliocentric phases in Table~4 were  computed according to the final
adopted period of pulsation and the epoch of CM Leo (P=0.$^d$361699 and
E=2450841.362285).   The resulting radial velocity curve is shown in Figure~8:
it has amplitude of 26.55 km s$^{-1}$, and the typical shape expected for a
{\it c} type RR Lyr. The systemic velocity ($\gamma$) of CM Leo was calculated
by integration  of the radial velocity curve on the full pulsation cycle and
corresponds to  24.47 km s$^{-1}$.

\begin{figure}
\vspace{7cm}
\includegraphics{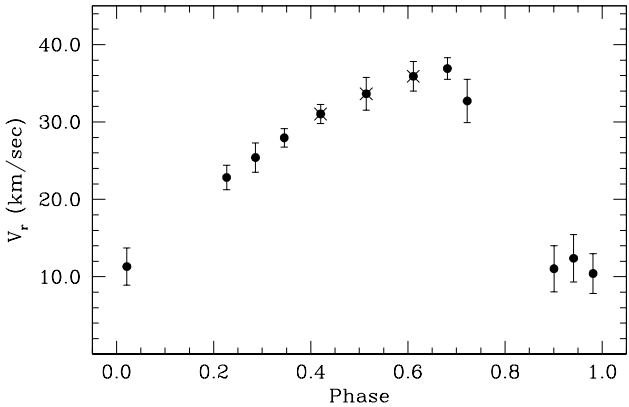}
\medskip
\caption{Radial velocity curve of CM Leo. The crosses indicates the three spectra 
taken around minimum light, and used to measure abundances (see Section 4.2).}
\end{figure}

\subsection{The metallicity of CM Leo}

Three spectra of CM Leo taken at/near minimum light (0.420 $ < \phi <$ 0.611) 
were used to measure the metal abundance of the variable. They were aligned to
the same radial velocity, then summed: the coadded spectrum has signal to noise
ratio at order centres of about 50 to 100 [namely: about 40 at 8400, 90 at 6000
and 5000, and 50 at 4000 \AA]. We derived temperatures T$_{eff}$ from the
dereddened  $B-V$ and $V-I$ colours of CM Leo at minimum light (with
$E(B-V)$=0.031 mag,  and $E(V-I)$=0.040 mag from Schlegel, Finkbeiner \& Davis 
1998 reddening maps).  Table~5 lists the dereddened $B-V$ and $V-I$ photometric
colours corresponding to the three exposures of CM Leo close to minimum light, 
along with the corresponding temperatures  as estimated  from Kurucz (1993a,b)
model atmosphere at [Fe/H]=$-$2.13 and $\log g$ =3.00.

\begin{table*}
\begin{center}
\caption{Temperatures derived from CM Leo dereddened colours near minimum light.}
\vspace{5 mm}
\begin{tabular}{c c c c c c c}
\hline
\hline
\multicolumn{1}{c}{Spectrum} &
\multicolumn{1}{c}{HJD(mid exp.)} &
\multicolumn{1}{c}{$\Phi$}&
\multicolumn{1}{c}{$(B-V)_0$} &
\multicolumn{1}{c}{$(V-I)_0$}&
\multicolumn{1}{c}{T$_{\rm eff}$}&
\multicolumn{1}{c}{T$_{\rm eff}$}\\
\multicolumn{1}{c}{} &
\multicolumn{1}{c}{}&
\multicolumn{1}{c}{} &
\multicolumn{1}{c}{}&
\multicolumn{1}{c}{}&
\multicolumn{1}{c}{$(B-V)_0$}&
\multicolumn{1}{c}{$(V-I)_0$}\\
\hline
W4 & 2451223.830009 & 0.420 & 0.353 & 0.396 & 6430 & 6878\\
W5 & 2451223.864014 & 0.514 & 0.357 & 0.409 & 6405 & 6843\\
W10 & 2451224.984161 & 0.611 & 0.338 & 0.403 & 6515 & 6858\\
\hline
\end{tabular}
\vskip 0.5 cm
\end{center}
\label{t:tab5}
\end{table*}

The average {\it photometric} temperature of the three spectra is T$_{\rm
eff}=6582$~ $\pm$ 250~K using the colour-temperature calibration and procedure
of Clementini et al.  (1995a)\footnote{These authors found that   Kurucz
(1993a,b) synthetic colours systematically overestimate stellar  temperatures.
This effect is larger for the $B-V$ and $V-I$ colours while almost negligible
for the $V-K$ colour-temperature calibration which is also  less metallicity
dependent (Fernley 1989 and references therein). In order to correct for  this
effect  following Clementini et al. (1995a) procedure (see their par. 3.4.2) 
the mean temperature derived from individual  $(B-V)_0$'s was lowered by 49 K,
and that from the $(V-I)_0$'s was  lowered by 97 K, before averaging them
together, to tie them to the  $(V-K)$-temperature calibration for RR Lyrae
stars.}. In principle, as an alternative approach, temperatures can also be
estimated by fitting the profile of Balmer lines. However, comparisons
performed in Paper II show that the temperatures derived in this way are within
100 K from the photometric ones. Since the present approach allows a direct
comparison with previous results on the same scale of Clementini et al (1995a)
and by-passes problems arising from the determination of the continuum region
near e.g. H$\beta$ (see Paper II), we prefer  to adopt the photometric
temperature, as given above.

We first measured equivalent  widths (EWs) for a number  of clean lines on the
coadded spectrum, using special purpose routines (Gratton 1988). Final EWs were
measured adopting a relation between EW and FWHM from clean lines and a
Gaussian fitting routine with only one free parameter, as fully explained in
Bragaglia et al (2001b).

Table~6 gives the linelist, the assumed $\log gf$'s and the measured  EWs. 

\begin{table}
\begin{center}
\caption{Linelist and measured EW's}
\vspace{5 mm}
\begin{tabular}{l c c r r}
\hline
\hline
\multicolumn{1}{l}{Element}&
\multicolumn{1}{c}{$\lambda$}&
\multicolumn{1}{c}{E.P.}&
\multicolumn{1}{r}{$\log gf$}&
\multicolumn{1}{r}{EW}\\
\hline
Fe I  &  4005.24 &  1.56 &  --0.57~ &   70.6      \\
Fe I  &  4063.59 &  1.56 &  --0.08~ &   98.1      \\
Fe I  &  4071.74 &  1.61 &  --0.02~ &   97.5      \\
Fe I  &  4132.06 &  1.61 &  --0.63~ &   65.0      \\
Fe I  &  4143.87 &  1.56 &  --0.44~ &   83.3      \\
Fe I  &  4235.95 &  2.43 &  --0.34~ &   42.8      \\
Fe I  &  4271.77 &  1.49 &  --0.16~ &   97.9      \\  
Fe I  &  4375.94 &  0.00 &  --3.03~ &   35.0      \\  
Fe I  &  4415.13 &  1.61 &  --0.61~ &   73.3      \\
Fe I  &  4427.32 &  0.05 &  --3.04~ &   18.9      \\
Fe I  &  4447.73 &  2.22 &  --1.34~ &   17.7      \\
\hline
\end{tabular}
\end{center}
Note - Table~6 is presented in its  entirety at the CDS. A portion is shown
here for guidance regarding its form and  content.
\end{table}

Standard spectroscopic abundance constraints (abundance results  from Fe~I
lines which show no trends with either expected line  strength or excitation or
ionization state), were easily satisfied in the analysis, confirming the
adopted photometric temperature  and surface gravity, and providing the
microturbulent velocity. We obtained average abundances of [Fe/H]=$-1.94$
($\sigma$=0.12 dex) from 21 Fe~I lines,  and [Fe/H]=$-1.92$ ($\sigma$=0.21 dex)
from 10  Fe~II lines, using a stellar atmosphere model with the following
parameters: T$_{\rm eff}=6582$~K, a surface gravity of $\log{g}=3.0$, and a
microturbulent velocity of $V_t=1.75$~km s$^{-1}$. 

Figure~9 shows the comparison of the spectrum of CM Leo, near the Mg b lines,
to the spectrum of CU Com, a double mode RR Lyr observed on the same run 
(Paper II), with [Fe/H]=$-$2.38, and whose lines appear, indeed, shallower. 

\begin{figure*}
\vspace{10cm}
\includegraphics{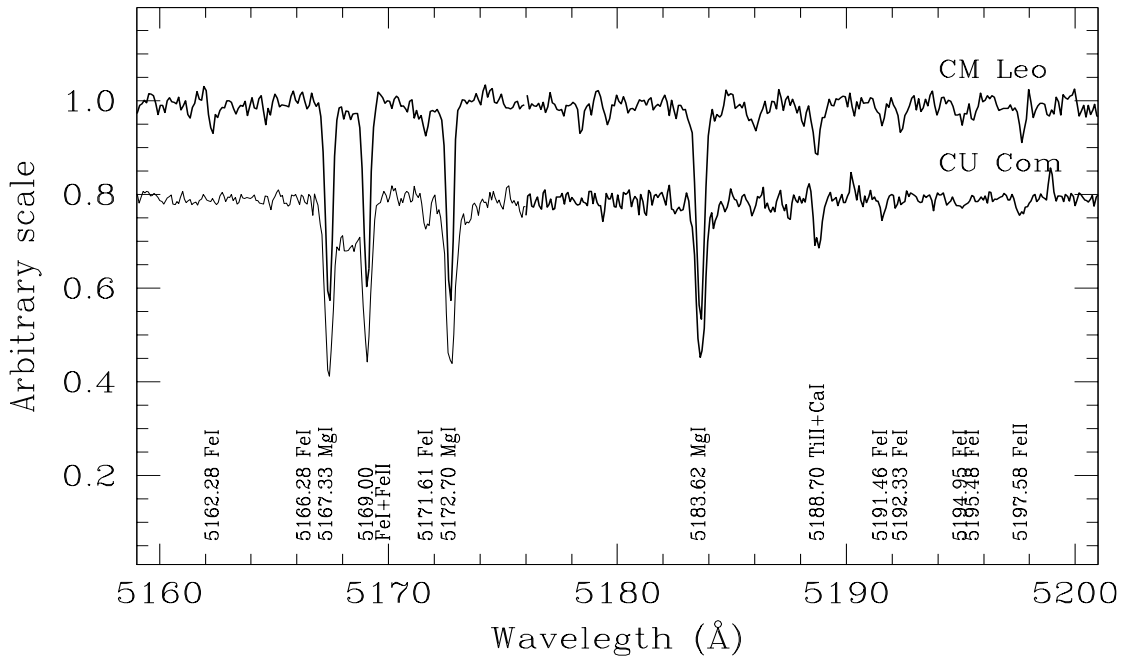}
\medskip
\caption{A portion of the added, rectified spectrum of 
CM Leo. Also shown for comparison is CU Com, for which Clementini et al.
(2000) measured [Fe/H]=$-$2.38.}
\end{figure*}

Uncertainties in the derived abundances are mainly due to possible errors in
the atmospheric parameters ($\pm$ 100 K in T$_{\rm eff}$, $\pm$ 0.3 dex in
$\log{g}$, $\pm$ 0.5 km s$^{-1}$ in V$_t$, and $\pm$0.2 dex in [A/H]) and on
the adopted model atmospheres (Kurucz 1993a,b).  Our estimate of the random
error contribution (including errors in measuring individual lines) is 0.15
dex. Thus, our conservative estimate of the metallicity and of the total error 
for CM Leo  is [Fe/H]=$-1.93\pm 0.20$. This metallicity is on the same scale of
Carretta \& Gratton (1997) and Clementini et al. (1995a) for RR Lyrae
variables. Table~7 shows the abundance ratios with respect to iron derived for
a  few other elements.
O, Na, and Al abundances include corrections for departure from LTE, and Mn and
Ba abundances include effects of the hyperfine structure. We find 
a large deficiency of Mn and Ba with respect to iron,
and an 
overabundance of $\alpha$-elements ranging from about 0.3 up to 0.8 dex 
for oxygen which appears to be particularly overabundant in this star.
We point out that large oxygen overabundances were found by Clementini et al.
(1995a) for three other RR Lyrae stars with [Fe/H]$< -1$ dex.

\begin{table}
\begin{center}
\caption{Average elemental abundances for CM Leo}
\vspace{5 mm}
\begin{tabular}{lcrcr}
\hline
\hline
Element    & (log n)$_\odot$ & Nr  & ${\rm[Fe/H]}$ &$\sigma$ \\
\hline
\hline
FeI	   & 7.50  &  21 & -1.94  & 0.12\\
FeII	   & 7.44  &  10 & -1.92  & 0.21\\
\hline
Element    & (log n)$_\odot$ & Nr  & ${\rm [el/Fe]}$ &$\sigma$ \\
\hline
${\rm [O/Fe]I	}$ & 8.73  &   2 &  0.82  & 0.03\\
${\rm [Na/Fe]I  }$ & 6.23  &   1 & -0.13  & 0.12\\
${\rm [Mg/Fe]I  }$ & 7.48  &   4 &  0.58  & 0.09\\
${\rm [Al/Fe]I  }$ & 6.23  &   2 &  0.42  & 0.12\\
${\rm [Ca/Fe]I  }$ & 6.18  &   4 &  0.40  & 0.19\\
${\rm [Sc/Fe]II }$ & 3.10  &   5 &  0.41  & 0.26\\
${\rm [Ti/Fe]II }$ & 5.14  &  11 &  0.36  & 0.23\\
${\rm [Cr/Fe]II }$ & 5.63  &   3 &  0.13  & 0.10\\
${\rm [Mn/Fe]I  }$ & 5.29  &   3 & -0.40  & 0.20\\
${\rm [Sr/Fe]II }$ & 2.88  &   2 &  0.34  & 0.07\\
${\rm [Ba/Fe]II }$ & 2.34  &   2 & -0.71  & 0.19\\
\hline
\end{tabular}
\label{t:elements}
\end{center}
\end{table}

\section{The modeling of the light and radial velocity curves}

We have compared the observed periodic  variations of light and radial velocity
with the theoretical predictions of models based on the solution of the
nonlinear hydrodynamical equations, including a non local and time dependent
treatment of convective transfer (Bono et al. 1997, Bono, Castellani \&
Marconi  2000, hereinafter BCM). These models not only reproduce the complete 
topology of the RR Lyrae instability strip but also predict the detailed
morphology of light and radial velocity curves.
As recently shown by  BCM in the case of the RRc U Comae,  the predictive
capability of current models has been tested by  reproducing the luminosity
variations of the pulsator along a full pulsation cycle. This approach, due to
the sensitivity of the predicted period and light curve morphology to stellar
parameters, supplies tight constraints on stellar mass, effective temperature,
intrinsic luminosity and, in turn, distance modulus.
To this aim we have applied a similar analysis to the observed  multiband 
light curve and radial velocity curve of CM Leo. We built pulsation models
based on the same physical and numerical assumptions as in  BCM but for the
observed metal abundance Z=0.0002 and an assumed helium  content Y=0.24. The
input mass, luminosity and effective temperature were varied in order to
simultaneously reproduce the observed  period (P=0.$^d$3617), the $V$ band
amplitude (A$_V \simeq$0.5 mag) and light  curve  morphology.
As extensively discussed by Bono \& Stellingwerf (1994), Bono, Caputo,
Castellani \& Marconi (1997), at fixed metallicity the details of predicted
light variations along a pulsation cycle mainly depend on the adopted
luminosity and effective temperature with a marginal sensitivity to the input
stellar mass. In particular for Z in the range 0.0001-0.001 the double maximum,
almost sinusoidal shape characterizing both CM Leo and U Comae, is typical of
hot (6800 K $\le T_{eff}  \le$ 7100 K) high luminosity (1.7 $\le
\log{L/L_{\odot}} \le 1.9$)  first overtone models. Moreover the amplitude of
first overtone models increases with the stellar mass, at fixed effective
temperature (see also BCM). 
Finally, as simply inferred by Ritter's
equation between the pulsation period $P$ and the mean density $\rho$,
combined with Stephan-Boltzman's law, and more accurately predicted by
nonlinear convective pulsation models (Bono et al. 1997), the period is
correlated with the luminosity and anticorrelated with both the effective
temperature and the stellar mass.
 All these dependencies
significantly restrict the parameter space and allow  us to single out the 
model that best reproduces the observed behaviour of CM Leo. Such a best fit
model is overplotted to the observed $V$ band light curve in  Figure~10 (top
right panel). 

\begin{figure*}
\vspace{16cm}
\includegraphics{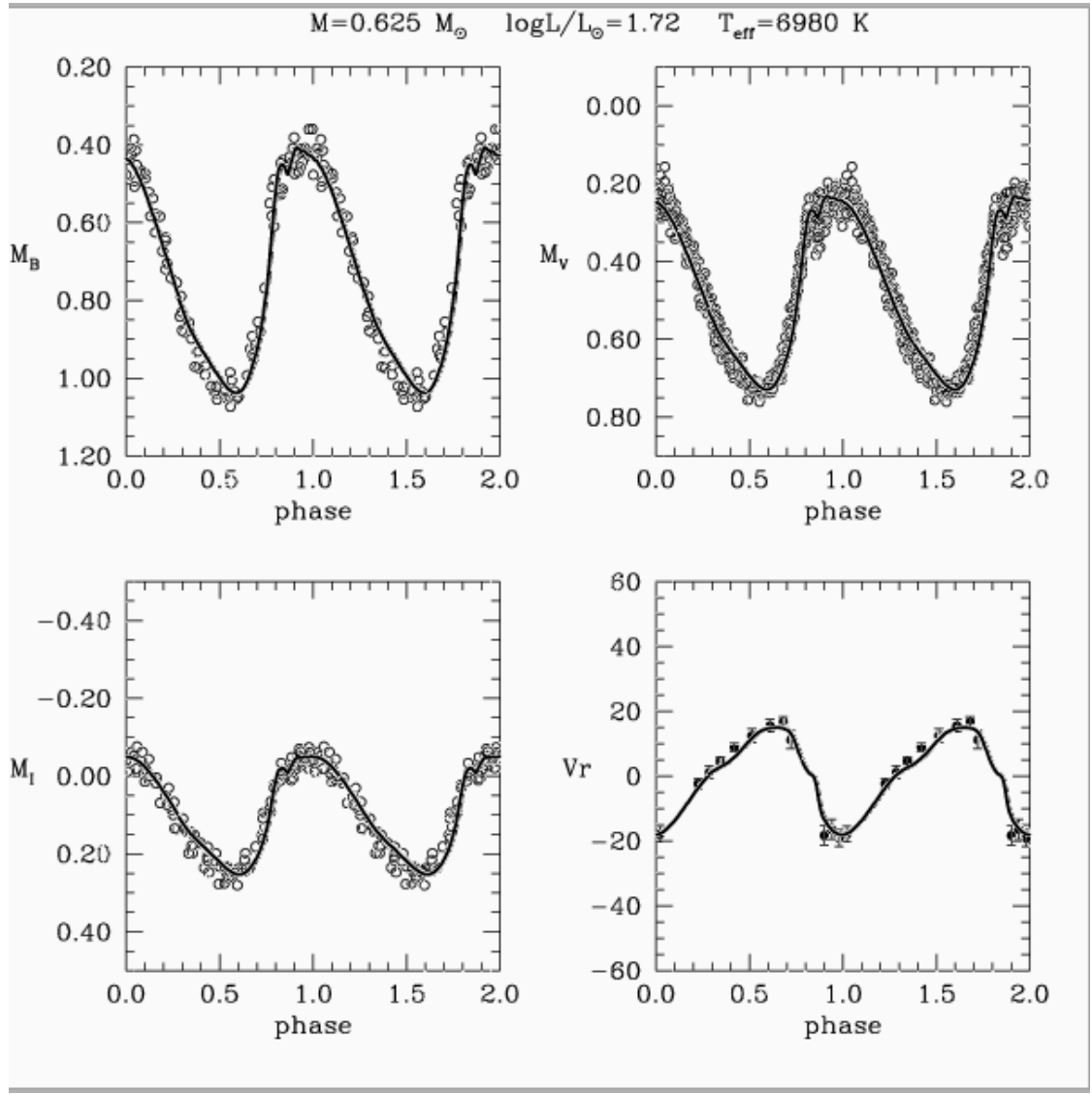}
\medskip
\caption{The multiband light curves and the radial velocity curve of the 
best fit pulsational model overimposed to the data.}
\end{figure*}

The intrinsic stellar parameters are labeled and the  corresponding  apparent
distance modulus is $\mu_V=13.20$ mag. By assuming  $E(B-V)$=0.03 mag and
$A_V=3.1 E(B-V)$ (Schlegel et al. 1998), we obtain a true distance modulus 
$\mu_0=13.11$ mag. The same model also reproduces the $B$ (top left panel)  and
$I$ (bottom left panel) light curves for the same $\mu_0$, at least within  the
photometric uncertainties ($\sim 0.03$). The agreement in different bands 
testifies that the colours of the best fit model well reproduce the intrinsic 
colours of CM Leo.
The bottom right panel of Figure~10 shows the comparison between the  observed
and predicted radial velocity curves when a baricenter velocity of  24.47 Km/s
velocity and a projection factor\footnote{The  projection factor is the
conversion term from radial to effective pulsation velocity, taking into
account the geometrical projection of stellar velocity fields corrected for
limb darkening effects.} of 1.36 are adopted.
The agreement is remarkable within the uncertainties on individual data points.
As discussed by BCM, the morphology of radial velocity curves is more 
sensitive than the light variations to the details of the coupling between 
dynamical and convective motions within a pulsating star.  Therefore the
agreement  found in Figure~10 confirms the high predictive capability of
current  pulsation theory.

\section{Discussion and conclusions}

CM Leo is a very regular first overtone RR Lyrae (RRc) with  period
P=0.$^d$361699, epoch E=2450841.$^d$362285,  a prominent hump  on the raising
branch of its light curves, and metal abundance [Fe/H]=$-$1.93 $\pm 0.20$.
Final results  as well as the average quantities derived for the star  from our
photometric, pulsational and spectroscopic analyses  are summarized in
Table~8. 

We have shown that the irregularities found by Clementini et al. (1995b) in
the  light curve of CM Leo disappear when star C3 is used as comparison star, 
since they were in fact produced by variations of about 0.1 mag  in the light
of the reference star used in Paper I (star C1). 
The true distance modulus   $\mu_0=13.11 \pm$ 0.02 mag we derive for CM Leo
leads to  a distance from the Galactic  disc\footnote{The distance from the
galactic plane almost coincides  with the distance from the Sun since CM Leo is
located in the  direction of the Galactic pole.} of $d=4.2$ kpc. This distance
is fully   consistent with the rather low metallicity we found for the star,
thus removing  the previously reported metallicity-distance anomaly of CM Leo.
We recall that a similar  result was found also for CU Com (see Paper II). 

RR Lyrae stars are known to be powerful standard candles through their 
absolute magnitude that is only slightly dependent on metal abundance. A {\it
pulsational} estimate  of the absolute magnitude of an RR Lyrae star provides
an independent  calibration of these standard candles and can thus be used  to
infer the  distance of any other stellar system where RR Lyrae's are found, 
the  Large Magellanic Cloud in particular. The absolute magnitude we derive for
CM Leo  from the  modeling of the multicolour light and radial velocity  curves
is M$_V$=0.47$\pm$0.04 mag, where the error includes the random uncertainty 
contributions
of the photometry (0.03 mag), and of 
the fitting with the theoretical pulsational models (0.02 mag in distance 
modulus and 0.01 mag in $\log{L}$, respectively).

\begin{table}
\begin{center}
\caption{Properties of CM Leo}
\vspace{5 mm}
\begin{tabular}{c c}
\hline
\hline
Type  & RRc \\
${\rm [Fe/H]}$& $-1.93 \pm 0.20$\\
Epoch & 2450841.$^d$362285 \\
P & 0.$^d$361699$\pm$ 0.000001 \\
$< V >$ & 13.66\\
$< B >$ & 13.97\\
$< I >$ & 13.32\\
$< B > - < V >$ & 0.31\\
$< V > - < I >$ & 0.34\\
$A_V$ & 0.498\\
$A_B$ & 0.631\\
$A_I$ & 0.301\\
$A_{\rm RV}$ & ~~26.55 km s$^{-1}$\\
$\gamma$ & $-$24.47 km s$^{-1}$\\
M$_V$ & 0.47\\
$\mu_0(V)$ & 13.11\\
\hline
\end{tabular}
\end{center}
\label{t:tab6}
\end{table}

Allowing for an evolution  of about 0.08 mag off the Zero Age  Horizontal
Branch for the RR Lyrae stars, as suggested by  Caputo \& Degl'Innocenti
(1995), and Caloi, D'Antona  \& Mazzitelli (1997), and  a slope of 0.2 mag/dex
for the luminosity-metallicity relation of  RR Lyrae's to correct CM Leo
absolute magnitude to the   average metal abundance of the RR Lyrae in the
Large Magellanic Cloud  ([Fe/H]=$-$1.50: Alcock et al. 1996, Bragaglia et al.
2001a),  the absolute magnitude we derive for CM Leo implies a value for  the
true distance modulus of the LMC of  $\mu_0$(LMC) = 18.43$\pm$0.06 mag,  
for an average 
dereddened luminosity of the RR Lyrae in the LMC bar $<V_0(RR)>$=19.07$\pm$0.05
(Clementini et al. 2001),  in better agreement with the {\it long} astronomical
distance scale.

\bigskip\noindent

ACKNOWLEDGEMENTS

This paper is based on observations  obtained with the 1.52 m telescope of the
Bologna Observatory in Loiano, the Michigan State University 60 cm telescope,
and the 2.7 m telescope of the  McDonald Observatory.
This work was partially funded by MURST-Cofin00 under the project
``Stellar Observables of Cosmological Relevance". 
SDT was partially supported with an Italian CNAA fellowship 
at the University of Texas at Austin. She thanks all the Texas Staff for the
warm hospitality. III gratefully acknowledges support from Continuing 
Fellowships at the University of Texas at Austin. HAS thanks C. Wilkinson for 
assistance in obtaining the 1998 MSU
observations and thanks the US National Science Foundation
for support under grant AST9986943. 
CS is happy to acknowledge that this research was partially funded by NSF 
grants, most recently AST9987162.

We warmly thank Raffaele Gratton for interesting suggestions and support.


\begin{thebibliography}{}

\bibitem{} Alcock et al. ({\it The MACHO collaboration}) 1996, AJ, 111, 1146

\bibitem{} Barning, F.J.M. 1963, Bull. Astron. Inst. Netherlands, 17, 22

\bibitem{} Blazhko, S. 1907, Astron. Nachr., 175, 325

\bibitem{} Bono, G., Caputo, F., Castellani, V.,  Marconi M. 1997, A\&AS, 
 121, 327

\bibitem{} Bono, G.,  Stellingwerf, R.F. 1994, ApJS, 93, 233

\bibitem{} Bono, G., Castellani, V., Marconi M. 2000, ApJ, 532, L129 (BCM)

\bibitem{} Bragaglia, A., Gratton, R.G., Carretta, E., Clementini, G.,  Di 
Fabrizio, L. 2001a, AJ, 122, 219

\bibitem{} Bragaglia, A., et al. 2001b, AJ, 121, 327

\bibitem{} Caloi, V., D'Antona, F., Mazzitelli, I. 1997, A\&A, 320, 823

\bibitem{} Caputo, F., Degl'Innocenti, S. 1995, A\&A, 298, 833

\bibitem{} Carretta, E., Gratton R.G. 1997, A\&AS, 121, 95

\bibitem{} Castellani, V., Maceroni, C., Tosi, M. 1983, A\&A, 128, 64 

\bibitem{} Clementini, G., Carretta, E., Gratton, R.G., Merighi, R., Mould, 
 J.R., McCarthy, J.K. 1995a, AJ, 110, 2319

\bibitem{} Clementini, G., Tosi, M., Bragaglia, A., Merighi, R.,  Maceroni,
 C. 1995b, MNRAS, 275, 929 (Paper I)

\bibitem{} Clementini, G., Gratton, R.G., Bragaglia, A., Carretta, E., Di
  Fabrizio, L., Maio, M., 2001, AJ, submitted, astro-ph/0007471 

\bibitem{} Clementini, G., Di Tomaso, S., Di Fabrizio, L., Bragaglia, A.,
Merighi, R.,  Tosi, M., Carretta, E., Gratton, R.G., Ivans, I.I., Kinard, A.,
Marconi, M., Smith, H.A.,  Wilhelm, R., Woodruff, T., Sneden, C. 2000, AJ,
120, 2054 (Paper II)

\bibitem{} Di Fabrizio, L. 1998, Laurea thesis, Univ. Bologna

\bibitem{} Fernley, J.A. 1989, MNRAS, 239, 905

\bibitem{} Gratton, R.G. 1988, Rome Obs. Prepr. Ser. 29

\bibitem{} Kholopov, P.N., et al. 1985, {\it General Catalog of Variable 
Stars}, IV ed. (Moscow; GCVS4)

\bibitem{} Kurucz, R.L. 1993a, CD-ROM 13, ATLAS9 Stellar Atmosphere Programs
and 2 km/s Grid (Cambridge: Smithsonian Astrophys. Obs.)

\bibitem{} Kurucz, R.L. 1993b, CD-ROM 18, SYNTHE Spectrum Synthesys Progams and
Line Data  (Cambridge: Smithsonian Astrophys. Obs.)

\bibitem{} Landolt, A.U. 1983, AJ, 88, 439

\bibitem{} Landolt, A.U. 1992, AJ, 104, 340

\bibitem{} Lomb, N.R. 1976, Ap\&SS, 39, 447

\bibitem{} Olech, A,. Kaluzny, J., Thompson, I.B., Pych, W., Krzeminski, W.
 Shwarzenberg-Czerny, A. 1999, AJ, 118, 442 

\bibitem{} Scargle, J.D. 1982, ApJ, 263, 835

\bibitem{} Schlegel, D.J., Finkbeiner, D.P., Davis, M. 1998, ApJ, 500,
525

\bibitem{} Tull, R.G., MacQueen, P.J., Sneden, C., Lambert, D.L. 1995, 
PASP, 107, 251


\bibitem{} Wilson, R.F. 1953, {\it General Catalogue of Stellar radial 
velocities}, Carnegie Institution of Washington Publ. 601, Washington D.C.



\end{thebibliography}
\end{document}